# Reproducing and Extending Experiments in Behavioral Strategy with Large Language Models

Daniel Albert
Drexel University
da692@drexel.edu

Stephan Billinger
University of Southern Denmark
sbi@sam.sdu.dk

In this study, we propose LLM agents as a novel approach in behavioral strategy research, complementing simulations and laboratory experiments to advance our understanding of cognitive processes in decision-making. Specifically, we reproduce a human laboratory experiment in behavioral strategy using large language model (LLM) generated agents and investigate how LLM agents compare to observed human behavior. Our results show that LLM agents effectively reproduce search behavior and decision-making comparable to humans. Extending our experiment, we analyze LLM agents' simulated "thoughts," discovering that more forward-looking thoughts correlate with favoring exploitation over exploration to maximize wealth. We show how this new approach can be leveraged in behavioral strategy research and address limitations.

Acknowledgements: We thank participants of the Strategic Organization Design Seminar (SOD) for helpful comments on a prior version of this research.

# 1 Introduction

A central interest of behavioral strategy research focuses on how cognitive processes and mental representations influence decision-making (Gavetti and Rivkin 2007, Levinthal 2011, Simon 1947). Two prominent approaches have emerged to advance our understanding of these microfoundations of strategy: computational work and human lab experiments. Agent-based computational simulations have sharpened our understanding of performance and learning consequences stemming from differences in individuals' cognition (Csaszar and Levinthal 2016, Gavetti and Levinthal 2000, Knudsen and Srikanth 2014, Winter et al. 2007). Additionally, scholars have increasingly designed experiments to study human responses within various tasks, such as searching for high-performing alternatives in unknown decision-spaces (Bergenholtz et al. 2023, Billinger et al. 2014, 2021, Richter et al. 2023), self-selecting into specific organizational tasks (Raveendran et al. 2022), exhibiting organizational voting behavior (Piezunka and Schilke 2023), and making innovation choices in response to different organizational contingencies (Klingebiel 2022).

Despite significant strides, a key challenge in advancing behavioral strategy lies in building and testing theories of individual-level cognition and its effects on the revealed decisions that our field typically focuses on. More theoretical development and empirical testing are needed to understand when and why decision-makers follow particular heuristics in specific situations, and what task factors influence their cognitive processes. Without addressing these questions, the field remains limited in its ability to explain effective decision-making. Important steps in this direction have been made in studies investigating strategic intelligence (Levine et al. 2017), emotions and strategic decision-making (Meissner et al. 2021), and the analysis of think-aloud experiments to uncover cognitive processes at play (Laureiro-Martinez et al. 2023, Reypens and Levine 2018). However, the resources and specialized experimental designs required for such studies can be challenging and thus limit the breadth and depth of research in this area.

In this paper, we explore how generative artificial intelligence, particularly large language models (LLMs), can advance our understanding of behavioral strategy. We build on Newell and Simon's efforts to use machines as proxies for studying human decision-making by "simulating human thinking" (1959, 1961). LLMs have demonstrated remarkable problem-solving capabilities across general and specialized domains. Recent research indicates that their outputs can closely resemble specific human behaviors in areas such as market research (Li et al. 2024), economic behavior (Mei et al. 2024), and social behavior (Ashokkumar et al. 2024). Given their growing ability to mimic human-like reasoning, LLMs offer significant potential for advancing research on cognition and decision-making in the behavioral strategy field.

Our objective is to evaluate the potential and limitations of LLMs as simulated participants in experimental research in behavioral strategy. To achieve this, we build on the experimental design of the "alien game" (Billinger et al. 2021) using LLM agents to generate responses similar to human participants. This experiment is designed to study how participants search a complex decision landscape, when performance-consequences and interactions between decision attributes are unknown – a prominent task and challenge of strategic management (Leiblein et al. 2018, Porter and Siggelkow 2008, Steen 2017). We compare our LLM produced results with those from the original human experiments to assess the effectiveness of this approach. We find that the LLM is remarkably effective in reproducing human search behavior and strategic decision-making, finding comparable results to revealed human behavior in this experiment. In addition, we extend the experiment by analyzing LLM agents' simulated "thoughts" concerning forward looking and backward looking (Gavetti and Levinthal 2000). We find that increased forward-looking "LLM cognition" is associated not only with greater search distance but also with a higher likelihood of ceasing search. This occurs as the agent weighs consequences of immediate benefits of maximizing known income (exploitation) against the potential of discovering better alternatives (exploration).

A core limitation we find, consistent with recent research on LLMs in other domains, is that LLM-produced results show lower variance than human samples. However, we propose and demonstrate that this reduced variance can potentially be mitigated by sampling from multiple LLM model "populations." By including a proportion of participants from a less capable LLM model, we achieve results that closely match the search behavior and variance of the original human sample. Given our findings, we argue that LLMs can constitute a novel and complementary path in advancing behavioral strategy research.

The remainder of the paper is organized as follows: We start by providing the conceptual background to our study, followed by a description and explanation of how we adjust a traditional lab experimental design to make it usable to "LLM agents." We then discuss our findings, followed by a discussion and conclusion.

## 2 Background

### 2.1 The beginnings of thinking-machines and research on strategic problem-solving

More than seven decades ago, Allen Newell and Herbert Simon developed a computer algorithm they termed the General Problem Solver (GPS), a program able to tackle complex tasks typically reserved for human intelligence, such as chess and military decision-making. However, the motivation to develop the GPS was not to replace human intelligence but instead to study it. They referred to this technique as "simulating human thinking" (Newell and Simon 1961). Newell and Simon conjectured that if one can simulate basic human cognitive processes, one may use these simulations to systematically study and build theory on human decision-making. For example, by simulating what they learned from expert chess players, they were able to systematically study the contingencies and consequences of "chunking" of memories, a cognitive pattern that occurs in expert chess players which allows them to categorize a plethora of complex board positions into higher order chunks, which eases them to recall them more reliably when needed (Chase and Simon 1973). This line of research concluded that chess masters learn domain-specific templates, based on pattern recognition, that help them to quickly navigate and effectively perform in the game (Gobet and Simon 1996).

In the decades to come after the GPS, chess playing has prevailed as an important setting in strategic management to advance theories on complex learning and decision-making processes. While computers have outcompeted humans in this particular task for decades, it is interesting that AI is still not a substitute for human thinking in chess playing (Gaessler and Piezunka 2023), suggesting that more complex cognitive processes are needed to capture strategic thinking. Several scholars in strategy have started to explore these complex cognitive processes associated with decision-making, arguing that decision makers may rely on heuristics, such as rational deduction, local search, analogical reasoning, and mental experimentation (Farjoun 2008, Gavetti et al. 2005). In this context, the role of cognition has always been highlighted as an important component of strategy (Gavetti and Rivkin 2007) and its potential to be linked to outcomes (Kaplan 2011), which is essential to behavioral strategy.

**2.2 Studying behavioral strategy in the lab**

Over the last decade, experiments have become an important method for studying behavioral strategy by allowing for a focused examination of causality and questions relevant for strategic decision-making. One starting point of this movement was the realization that simulation modeling, which was and is highly prevalent in the field, often relies on numerous assumptions concerning human behavior that have only received limited empirically testing before. For example, search of computational agents is in most simulations modeled as local search with occasional long jumps. This conceptualization derived from previous literature on NK (Kauffman 1993) as well as insights on how organizations function (Cyert and March 1963, March and Simon 1958). Whether this assumption would hold in the laboratory was then first examined by Billinger et al. (2014) who allowed human agents to search rugged landscapes. One take-away from their study is that human agents, on average, search more broadly than what the strict local search assumptions would suggest (more below). This empirical finding was then confirmed by others who employed similar experiments (Bergenholtz et al. 2023, Richter et al. 2023, Tracy et al. 2017), and who were joined by others interested in the investigation of actual revealed behavior.

The simulation of human thinking as well as human lab experiments, however, have fallen somewhat short on simulating just that, the human thinking processes. Consequently, such models may study the consequences of a particular heuristic or mechanism, but not the reasoning or justification of "why" an agent draws in a particular situation on one heuristic or rather than another. One way to address this question is the study of human attention, which constitutes a key cognitive process that is central to strategic decision-making as it influences where and how decision-makers search and solve problems (Gavetti et al. 2012, Ocasio 1997). While significant progress was made in studying attention conceptually and empirically (Joseph et al. 2024), this line of inquiry has not put an emphasis on studying actual human thinking processes. The study of these processes has been addressed by research that builds on so-called "think aloud" protocols (Ericsson and Simon 1998), where participants in experiments are asked to articulate every thought they encounter while working on a particular task. What humans share in their think aloud protocol is not only supposed to describe where their attention goes during their decision-making process, but also serves as a basis for understanding cognitive flexibility (Laureiro-Martínez and Brusoni 2018) and how problem-solving strategies emerge and unfold (Laureiro-Martinez et al. 2023).

Despite its great relevance for the field of behavioral strategy, think-aloud protocols and other attempts to capture human cognition processes are challenging for at least two major reasons: First, administering think-aloud protocols, where the researcher asks the participant to share every thought that crosses their mind while working on the actual task (e.g., solving a problem) must be carefully applied to an experiment to not interfere with the participants behavior of interest (e.g., how they solve the problem). Ericsson and Simon (1998) dedicated much of their scholarly discussion on this challenge. More recently, researchers showed that think-aloud protocols can constitute a powerful component of behavioral experiments (Leighton 2017). However, to ensure that the think-aloud protocol is not altering the participants' behavior, counterfactual studies would often be useful but are rarely conducted because of the added complexity and cost. Second, think-aloud protocols can exert substantial resource strains on experiments as participants may not take certain decisions in an experiment at the same time as they are supposed to verbalize their thoughts, raising consistency challenges in administering the protocol. Despite these challenges, think-aloud protocols have uncovered important insights showing, for instance, how the manipulation of attention can shift a decision-maker's emphasis from framing to implementation during problem-solving (Laureiro-Martinez et al. 2023). Disentangling cognitive processes with think-aloud protocols holds great potential but remains difficult and costly to implement together with the risk of small deviations to jeopardize the experiment. However, the recent advances in generative large language models may allow complementing human experiments via the study and simulation of human thinking preceding or accompanying human lab experiments.

## 2.3 Large language models mimicking "human thinking"

Large language models (LLMs) are a specific class of generative AI models with the objective to generate human-like language outputs in response to language input – referred to as prompts. These LLMs have seen a particular rise in popularity with the launch of OpenAI's ChatGPT class models in 2022. This recent breakthrough in LLMs is in large part attributed to the "transformer" architecture (Vaswani 2017), which enables the LLM to interpret and build upon "context" in its responses. It contextualizes conceptual meaning – expressed as a hyper dimensional vector – to words and word fractions (so called tokens) that take on different values depending on the words (and their vector values) around them – referred to as the context window. These values are learned by the LLM as part of a massive training process that is largely unsupervised, that is, without human-assigned labeling of data. In this process, LLMs are provided with a vast amount of written content, where the AI trains itself to predict the next word (or token) by relying on a self-attention mechanism that considers the entire content provided (oversimplified: one may think of the LLM reading some of the words while covering and considering the rest of the sentence or paragraph). The learned values for a particular token reflect its mathematical position in the vector space.

For example, in a word vector space study, researchers found that the words man and woman have the same distance on a particular dimension as the words king and queen. They demonstrated that through vector arithmetic, the model could arrive at the word "queen" simply by calculating "king - man + woman" (Mikolov et al. 2013: 746). This "reasoning" through a vector space helps explain why language models are particularly capable of solving analogy problems and capturing semantic relationships between words. At the same time, the resulting output based on reasoning is probabilistic in nature, that is, the predicted next word depends on what the overall model it was trained with (typically large amounts of data) and the token probability distribution that was derived for each word. That is, the same context presented to the same LLM twice typically produces each time a new response that results from a distribution of contextualized word "vectors."[1] A repeated LLM prompting of the same input thus will result in a distribution of responses rather than a single type of response. Given these behaviors, LLMs can be considered to be "implicit computational models of humans", or a "homo silicus" (Horton 2023), as an analogy to the economists' "homo economicus", and thereby become the basis for new method developments in behavioral strategy.

---

[1] This probability distribution can be adjusted with a parameter called "temperature", which controls the degree of randomness. With a lower temperature (values close to zero), the distribution is more peaked and the model becomes more deterministic, resulting in quasi-identical outputs for the same input. With a high temperature (values close to the maximum 2), the distribution "flattens" and the model becomes more divergent, creating outputs, for the very same input, that show greater variation.

## 2.4 LLMs studying human cognition and behavior in different fields

Various academic fields have started to examine how well LLMs mimic human cognitive behavior and there are many promising results. Overall, the body of literature shows a common thread of LLM's remarkable effectiveness in mimicking many features of human cognitive behavior, while also highlighting their limitations. For instance, in economics, Mei et al. (2024) administered a Turing test to AI testbots and found that ChatGPT4 exhibits behavioral and personality traits that are statistically indistinguishable from a random human. They also found that in strategic situations, LLM behaviors tend to be more altruistic and cooperative than average human behavior. In political science, Bisbee et al. (2024) have examined if LLMs can replace human survey data in public opinion research. The results show that ChatGPT can reproduce averages, but not the variance found in human samples. In psychology, Strachan et al. (2024) showed that GPT-4 matches human performance in various theory of mind tasks, which test the ability to understand and infer others' beliefs and intentions, but, at times, shows more cautious decision-making than humans. In marketing, Li et al. (2024) have examined if LLMs can replace human surveys for perceptual analysis in market research, and they found that LLM-generated data closely aligns with human responses, achieving up to 87% agreement, but with reduced variability in some areas.

The abilities observed in LLMs have expanded with newer models having achieved responses that essentially resemble that of human responses and can take on expert roles in certain domains, such as PhD level mathematics (Franzen 2024). This broad level of human-like behavior and expertise along different contexts may pose great opportunities both for general and highly context-specific questions, such as strategic management. For example, Csaszar et al. (2024) examine a highly specialized context of evaluating business strategies and indeed find that LLM's can generate and evaluate strategies at a level comparable to entrepreneurs and investors. At the same time, Doshi et al. (2024), in another study that evaluates strategic decision-making and LLMs, find similar results for aggregate evaluations, but with the caveat that generative AI (comparing multiple LLMs) often produces evaluations that are inconsistent and biased.

In sum, many of the reviewed studies conclude that LLMs are often effectively mimicking human cognitive responses and can produce (average) outcomes that are indistinguishable from human responses. However, a common limitation noted in most studies is the lack of the full variability observed in humans.

## 3 Reproducing and Extending Experimental Designs in Strategic Management

In evaluating LLM's ability to mimic human behavior in ways that could prove to be relevant for future research in behavioral strategy, we need to specify an experimental context that meets criteria of tractability and availability of a solid body of prior work. A rich line of research has stressed that search for novel alternatives poses a fundamental task of strategists (Cyert and March 1963, March 1991). The decision-making associated with search is often portrayed as a "discovery" process where decisions interact in non-trivial ways (Leiblein et al. 2018, Porter & Siggelkow 2008, Van Steen 2018). A canonical framework in studying search is the NK fitness landscape, a multi-dimensional decision space that allows for varying degrees of performance interactions between decisions (Kauffman 1993, Levinthal 1997). The NK framework has informed behavioral strategy through the use of both agent-based simulations and human lab experiments which resemble a rich body of work that offers a particularly suitable context to explore LLM's simulated behavior. Next, we will provide a general introduction to the NK framework and its suitability to study questions of behavioral strategy under search.

### 3.1 Simulating complex problem spaces

The NK model was introduced to the field of management by Levinthal (1997) who demonstrated the framework's powerful suitability to study questions of organizational adaptation, the role of selection forces, and environmental change in the presence of epistatic interdependencies, that is, when the performance contribution value of one choice depends on its own state as well as on the states of K of the N-1 other choices. The NK model has seen broad adoption for questions core to the field of strategic management, such as, imitation of competitors (Csaszar and Siggelkow 2010, Rivkin 2000), industry evolution and profitability (Lenox et al. 2006), organizational design and exploration (Siggelkow and Rivkin 2006), and ecosystem innovation (Ganco et al. 2020).

Interest has emerged around questions of behavioral strategy using the NK model, with a focus on mental representations, cognition, and the consequences on search heuristics (Gavetti and Levinthal 2000, Winter et al 2007, Csaszar and Levinthal 2016). The suitability for this type of study using the NK framework stems from the underlying properties that are "tunable." Given a set of *N* decisions that can be configured in a variety of combinations (typically $2^N$ as for parsimony reasons decisions are fixed to binary states, such as 0 or 1, "on" or "off", "high" or "low", etc). Each of the N decisions contributes a performance value, where this value is dependent on the state of the focal decision (e.g., whether it is high or low) and the states of K other decisions. Thus, K denotes the level of complexity, with low values indicating little complexity and high values reflecting great levels of complexity. There are two corner cases: K = 0, where no dependencies between decisions exist, that is, each decision can be optimized without the need to consider any of the other decisions. As a consequence, the performance search space is often described as a "smooth," single peaked landscape, where the search for incremental improvements will eventually lead to the global optimum – i.e., the configuration where each decision is optimized to the highest of its two possible performance values.

The other corner case is K = N-1, where every decision is dependent on the states of all other decisions. In such a case, the landscape is simplistically described as "jagged," that is there are many local optima and a single decision change can fundamentally alter the performance as all decisions are affected by it. For non-zero values of K, scholars have described the search space as a rugged landscape, where local optima have basins of attraction, that is, once near a peak, incremental improvements will lead up to the local optimum – referred to as "hill-climbing." In other words, the performances of nearby positions in rugged landscapes are correlated but local optima within a landscape can vary substantially with respect to overall performance. While there are many mediocre peaks, some are particularly high in performance. Incremental search, however, proves cumbersome in such terrain as once attracted to a local basin, one would need to divert search to where performance is in fact lower, to traverse through a valley toward a potentially higher-performing area of the landscape. Because the landscape and its characteristics are not known to searchers but rather need to be discovered, the NK framework resembles a familiar conundrum of myopia and exploitation-exploration trade-offs.

**3.2 Bringing NK to the lab**

While the majority of NK research relies on simulation modeling (Baumann et al. 2019), lab experiments have investigated humans' revealed search behavior - devising the "Alien Game" to study how humans search rugged landscapes (e.g., Billinger et al. 2014, 2021, Bergenholtz et al. 2023, Richter et al. 2023), and whether and where they search (Billinger et al. 2021). The Alien

Game is an experimental setup that utilizes the NK engine as its backbone together with a front-end design that effectively allows participants to play a game of search on NK landscapes with a limited number of trials. Specifically, participants are instructed to envision they had made contact with a new species from out of space (the alien) and that they can design and sell art pictures. The art picture is made up of a combination of ten symbols that can be activated or deactivated, and that the alien's preference for a specific pattern of symbols is not known ex-ante but can be found out by selling such an art picture. This experimental design offers several advantages that serve multiple and important purposes for the study of behavioral strategy: First, by instructing participants about the unknown preferences of an unfamiliar species, this experiment is designed to reduce or remove any prior assumptions or experiences a participant may hold, thereby facilitating comparisons with computational agents. Second, the use of cryptic symbols further limited any wide-spread associations that may otherwise affect the search for combinations. Third, the game offers a limited yet not trivial set of strategy parameters, including the unambiguous measurement of feedback variables, varying task complexities and possibility to operationalize important exploitation-exploration trade-offs.

In a recent version of the Alien Game, these advantages are utilized by asking participants to maximize their accumulated income from the sale of art pictures (Billinger at al 2021). With this incentive scheme, participants effectively faced a search challenge with two distinct exploitation-exploration trade-offs: (1) *whether* and when to stop searching for better alternatives (i.e., "cashing in" on the best alternative found so far), and (2) *where* to search in the landscape by either narrowly searching in the neighborhood of existing solutions or searching more broadly in the landscape. The results of these experiments showed that different feedback variables, such as early feedback and immediate feedback, influence these two decisions differently, thereby extending our understanding of how individuals respond to feedback above/below aspiration and contributing to resolving inconsistencies identified within the problemistic search literature (Posen et al. 2018).

Overall, the Alien game's grounding in the well-studied NK framework, the validation of the experimental setup in multiple studies (Anvari et al. 2024, Tracy et al. 2017, Vuculescu 2017), and the possibility to investigate combinatorial multi-attribute decision-making make the alien game a suitable candidate to test how LLMs compare to humans when facing problem-solving behavior.

**3.3 Taking LLM to the lab**

We conducted our main experimental analysis using OpenAI's ChatGPT-4o (model 'gpt-4o-2024-08-06'), which was OpenAI's latest frontier model at the time. OpenAI is often considered to be a leader in offering advanced large language models (LLMs). The use of gpt-4o was particularly convenient due to its easy API-interface, allowing us to send and receive data via Python. This integration enabled seamless communication between our NK framework, which we programmed entirely in Python as a standalone local program, and OpenAI's servers. For the baseline comparison, we operated the LLM with Open AI's default settings.[2]

To assess the extent to which LLMs can simulate human behavior in the context of the alien game, we took multiple steps to reproduce the experimental design as close as possible to the original studies, making only such changes that are necessary to interface the experimental framework with the LLM.[3] We contacted the original author team to gain access to the published data and additional details on the experimental setup. As a result, we could rely on the same NK landscape used in the original study and use the same participant instructions to produce data that allows for a comparison of data from human agents with data from LLM agents. The changes that we needed to implement in order to make this experiment accessible to the LLM include converting the graphical computer interface that human participants saw as a screenshot in the original instructions into plain text. For example, instead of showing the ten cryptic symbols that participants could switch on or off, we named them along the Greek letters (alpha, beta, and so on) and the LLM had to indicate in a trial which of these symbols it wanted to switch on or off, naming them by the respective Greek letters. The conversion of the graphical interface into text also required that after each trial the LLM received a short reply on what the payoff of that round was and what the current overall wealth (accumulated payoffs) is. In the original experiments, participants would see this information on the screen together with all prior trials and payoffs. To ensure that the LLM would have a similar situation each trial, we set up the LLM to have all text input and output remain in the LLM's context history and in each trial, the LLM would need to be prompted to submit its next art picture configuration including choices for all ten Greek symbols. Specifically, we prompted "Considering what you know so far, please submit your next trial configuration."

---

[2] OpenAI allows varying parameters that affect token prediction. While this is interesting to study in itself, we chose to apply the default settings to gain a first baseline of insights.
[3] For example, we even replicated the requirement that a participant had to take a multiple-choice test on the experiment's instructions. That is, LLM agents were presented with a slightly adapted multiple choice test prior to playing the alien game.

  In the original study, the authors used a within-subject-design and asked each participant to play three separate blocks of the game with 24 trials each - one block for each of the three levels of landscape complexity with $K = 0$ (low), 5 (medium), and 9 (high). In the original experiment, participants were instructed that there was no correlation between the three blocks (since these represented dealing with three different aliens) and the participants were asked to treat these blocks as completely independent from one another. The original study, for that reason, included controls and robustness tests to rule out unwanted learning effects between blocks (Billinger et al. 2021). For practicality reasons[4], we ran each level of complexity as an isolated LLM experiment and thereby created a between-subject design, which, as a side-effect, also resolved possible between-landscape learning effects. Finally, the human experiment composed 69 participants, which is why we initially ran 69 independent runs for each landscape – that is, each run represents a single simulated LLM agent for that landscape.

## 4 Analysis

In this section, we start by comparing the LLM results with those from the human sample. Next, we extend the analysis by examining the LLM agents' generated thoughts and their relationship with outcome variables. In a third step, we explore the results obtained from sampling multiple LLM models. Finally, we discuss the robustness tests we conducted.

### 4.1 Reproducing a behavioral experiment with LLM

A main interest of the original alien game study pertains to "whether to search," i.e. the time participants take out of 24 trials to search actively for art picture configurations compared to when they tend to stop their search in an attempt to accumulate wealth - that is, selling their best picture for the rest of the trial periods. In Figure 1, we report LLM results (left plot) and the original human results (right plot) of the percentage of participants actively searching (y-axis) per trial period (x-axis), each landscape complexity is shown separately. The LLM results reproduce the overall trend of the original results, indicating a few periods of consistent search across all agents, followed by a relatively consistent decline in actively searching participants with an increase in trials. Notable is a similar level of active search at the end of the experiment of around 20 percent in trial 24 for both the LLM agents and the human participants. The LLM results, however, show greater variation between complexity levels, with LLM agents in very high complexity landscapes ($K = 9$) stopping their search somewhat earlier than their human counterparts. High complexity landscapes have greater performance variance even among neighboring search positions, which may disincentivize search in favor of wealth accumulation as a poor search trial may incur high opportunity cost.

---

[4] Because all text needs to be retained in the LLM's context window (i.e., the history chat), storing multiple game iterations may exceed the available context size and in addition increases processing costs, as the API incurs cost per tokens (i.e., input and outputs).

-------------------------------
Figure 1 about here
-------------------------------

Another interest of the original study was the question of "where" participants would search, that is, how many configurational changes they make, and thereby explore, relative to their currently best-performing art picture configuration they are aware of. Consequently, this search distance is measured as the number of configurational differences between a focal search trial and the participant's best prior configuration (also called Hamming distance; see Hamming 1950). In Figure 2, we report for actively searching participants, the search distance (y-axis) per trial (x-axis) in three panels, each of which is comparing human results and LLM results. The panel on the left shows these results for low complexity ($K = 0$), the center panel shows $K = 5$, and the right panel shows $K = 9$. The envelopes show the standard deviation for each graph, which allows to gauge variance in the data samples.

The LLM results generally reproduce the original results, that is, that participants early engage in more distant search, followed by more local search with an uptick in search distance toward late trials. However, the LLM overall search distance is generally lower than that of the human study. The search distance mean of the human sample is 2.36 and for the LLM sample 1.31 - the difference is significant with p<0.001. Upon visual inspection, we find that humans across levels of complexity tend to initiate their first search with an average of around four changes to the starting configuration. In contrast, LLM agents tend to only make around two changes. A similar difference can be observed towards the end, whereas in-between the beginning and the end of the experiment, human and LLM agents are much closer in search distances. A notable difference between the human and LLM results appear to lie in the observed variability. The human results show greater variability as reflected in the standard deviation envelopes that are substantially wider than for the LLM agents. (We return to and address this observation in 4.3.)

--------------------------------
Figure 2 about here
--------------------------------

To summarize the first set of results: Our analysis of the LLM results suggests a noteworthy reproduction of the human behavior reported previously (Billinger et al. 2021). Specifically, we find similar behavior with respect to how long participants search a landscape before they focus on maximizing their wealth within the remaining time. We also find a similar behavior with respect to where agents search, that is, initially starting broader, narrowing down search, with an uptick in search distance toward the end of the experiment. An important difference we find in all our LLM results is that search distance tends to be, on average, more local than in the human reported trials.

**4.2 Analysis of "thinking" patterns**

Considering our first results, which appear promising with respect to LLM agents reproducing human search behavior, we are also interested in the simulated "thoughts" of the LLM and whether an analysis of its content would provide theoretically useful insights. We therefore ran the same experiment again with a slight change to the prompt, asking the LLM to "think aloud." The observed results did not change between the main instructions (as reported under 4.1) and the additional encouragement to "think aloud." However, we found that the LLM provided slightly more thoughts in its responses when asked to think aloud in addition to our main instructions.

In this think-aloud extension of our LLM experiments, we focused our analysis on the role of cognitive attention as a key construct underlying the behavioral theory (Gavetti et al. 2012, Ocasio 1997). We are particularly interested in examining how the LLM-produced thoughts relate to the focus of attention in a given trial and how this attention relates to the search behavior. We draw on Gavetti and Levinthal's (2000) theory on the role of **backward looking** and **forward looking** decision-making. These authors have stressed that backward looking search is driven by trial-and-error experience, whereas forward looking decision-making is akin to a mental representation that is more predictive in nature.

**4.2.1 Measuring types of attention using LLM outputs**

Each trial period, the LLM provides a written response (output) in reference to our prompt to name the next trial configuration the LLM intends to test. We use this output as the simulated raw thoughts. From this written response, we extract text that relates to:

- ***backward looking***, that is, such statements that reference previous rounds and insights that have been learned in prior rounds.

- *forward looking*, that is, such statements that reference future trials, general strategy, and everything related to next steps.

Extracting and labeling statements as either backward or forward looking was done using a separate LLM API interface that we provided with the individual outputs of our original LLM agents after the experiments were completed. Utilizing LLMs for research tasks such as assigning labels and interpreting unstructured texts has shown great potential (Boussioux et al. 2023).

We measure attention by comparing the character counts of forward-looking statements to those of backward-looking statements (i.e., the ratio). This measure, albeit coarse, provides a proxy for the relative distribution of attention between past experiences and future predictions. The number of words used in human think-aloud protocols has frequently been employed to capture the direction of cognitive processes and their relative extent, as the quantity of verbalizations has been shown to reflect the focus of attention and the depth of cognitive engagement (Ericsson & Simon, 1993; Chi, 1997; Fox, Ericsson, & Best, 2011). While the LLM deploys no "cognitive processes" in a human sense, it mimics human reasoning and thought patterns expressed in language. We therefore assess how the LLM balances experiential learning with forward-thinking strategies expressed in its communicated "thoughts."

We also count the number of distinct symbols, out of the ten in the art picture configuration, that the LLM specifically mentions in its output. This unambiguous count measure, ranging from zero to ten, helps us capture the breadth of its attention (*attention breadth*). Note that this measure can differ from revealed behavior, where the LLM may only change one symbol but discusses three or four elements during deliberation.

### 4.2.1 Regression analysis of LLM outputs

Following Billinger et al. (2021), we also specify a 2-step Heckman regression model (Heckman 1979) to effectively capture the LLM agents' two stages of decision-making, that is, first whether to search (stopping or not), followed by where to search (search distance). We include relevant covariates from the original study, including initial feedback as the exclusion restriction for estimating the 1st stage (active search), along with our two attention variables. The regression output is included in Table 1 and a list of variable names and definitions is included in Table 2. For the think-aloud analysis, we created a sample of 900 independent LLM agents (i.e., 300 LLM agents for each of the three landscapes) to increase statistical power. Our regression models generally concur with Billinger et al. (2021), and we find similar relationships between the different feedback variables of interest (initial feedback, average feedback, and immediate feedback) and a participant's search distance and propensity to stop their search.[5]

-------------------------------
Table 1 about here
-------------------------------

For the new attention variables that we include in our study, we find that our variable Attention Breadth is associated with a lower probability of stopping search and greater search distance. That is, a LLM agent "thinking" about a greater number of symbols (i.e., decision attributes) is more likely to make a greater number of changes and continues to search. This finding for the mimicked thoughts of LLM agents may to some extent reflect prior findings of cognitive flexibility, which showed that more deliberate system-2 thinking would base problem-solving on more decision elements (Laureiro-Martínez and Brusoni 2018). Regarding the role of forward and backward-looking attention, we find greater forward looking attention (relative to backward looking attention) to be associated with a greater propensity of stopping but also a greater search distance. This is an interesting and unexpected finding as forward looking constitutes a variable that appears to induce exploration (i.e., "where to search"), while also planning ahead for the maximization of accumulated wealth, which requires to stop active search and "cash in" on a high alternative. Furthermore, the interaction term of trial number and relative forward-looking attention decreases the propensity of stopping. In other words, when getting closer to the end of the experiment, it appears that LLM agents are less likely to stop as there are only a few trials left, which may not make a substantial difference in maximization. If at that point no satisficing configuration has been found, the LLM agent appears to take chances and search – possibly hoping for a "big win." However, we find no indication that this is accompanied by a more aggressive effort to explore distant configurations as the interaction term is non-significant in the second stage model of the Heckman regression.

---

[5] We observe some deviations from the human sample results that are not central to our main findings. These differences may stem from changes in research design (e.g., agents playing one landscape instead of three). In addition, the original study had a significantly lower number of observations, for which reason the authors performed multicollinearity procedures (see Billinger 2021: 373). These procedures are not necessary for our analyses.

Figure 3 displays the relative level of forward-looking to backward-looking attention (measured by character count) across trials for the three landscape complexities. The three lines show a similar trend: forward-looking attention begins high, decreases, then increases again toward the experiment's end. This pattern reflects LLM agents starting without prior knowledge (high initial forward-looking deliberation, no backward looking), then engaging in experimental, backward-looking search (Gavetti and Levinthal 2000). The late increase in forward-looking attention reflects the LLM agents' planning of maximization of wealth, which requires careful consideration of whether to search (again) or accumulate the currently highest performance. This highlights the connection between the exploitation-exploration trade-off and forward-looking attention in both stopping and searching behaviors and thereby complements Gavetti and Levinthal's (2000) theory.

-------------------------------
Figure 3 about here
-------------------------------

**4.3 Increasing Variance by Mixing LLM Populations**

In our findings of 4.1, we showed that similar to other LLM studies, we too find lower variance in LLM responses compared to human samples. One argument that could be made is that we draw on a single "population" of LLM agents that are generated from the same model. Human samples in experimental studies are often also considered to originate from a specific type of population, such as university students (Raveendran et al. 2015, Richter et al. 2023). However, these populations typically vary substantially with respect to prior experiences, level and quality of education, nationality, and many other factors that can have an influence on participants' cognitive perception and decision-making. The human laboratory addresses this issue with randomization, which is not applicable in LLM experiments in a comparable way. However, this shortcoming could be addressed by using more than one LLM model and thereby relying on LLM agents from different LLM populations. Available LLM models differ in their training and abilities of inference from input, which one can distinguish into different classes of LLMs. There are frontier models, such as Meta's LLama 3.1 or OpenAI's GPT-4o, and there are also slightly less advanced models (with respect to the underlying architecture of parameters), such as Meta's LLama 2 or OpenAI's GPT-3.5.

To test the hypothesis that a LLM model mix can better mirror a human participant population, we use results from section 4.1 (from GPT 4o) and include an additional 20% (14 cases) of LLM agents stemming from one of the less advanced models, namely GPT 3.5 Turbo, in an attempt to reproduce a greater span of participant variability. This particular model has received substantial research attention and has been found to be mimicking human behavior also in remarkable resemblance but has been described as less capable as current frontier models.

In Figure 4, we report the comparison of the human experimental data with the mixed LLM population data, (69 case GPT 4o + 20% (14 cases) GPT 3.5). The mixed LLM population data now closely resembles the human trial data in values as well as variance. Specifically, the LLM mixed sample has a search distance mean of 2.20 with a standard deviation 1.93 which is very close to the human sample with a mean of 2.37 and standard deviation of 1.97. These findings suggest that the current limitations associated with reduced variability within models, may be overcome, depending on research objectives, by using multiple LLM models.

--------------------------------
Figure 4 about here
--------------------------------

**4.4 Robustness tests**

We examined how sensitive our results were to certain experimental implementation choices that were necessary to make the experiment accessible to our LLM-based research framework. First, we reproduced an earlier study by Billinger et al. (2014) that focuses on finding the highest possible configuration without any consideration of wealth maximization. We found similar alignment between our LLM agents and the human study as we did in the reproduction of Billinger et al. (2021). Second, we tested alternative framing designs, that is, instead of specifying the "alien game," we tested two alternatives. In one alternative, we described a setting of a new species of animals, discovered in an unknown bio-system on earth, with unknown nutrition. The LLM agents are then asked to test combinations (high and low levels) of ten nutrients. The other experimental framing was a technical "barebone" framing, where the LLM was instructed to consider combinatorial alternatives of '0' or '1' leading to different performance outcomes. For both framings, the results are qualitatively comparable.

Third, we tested many small deviations from our main instruction set to identify if the LLM was particularly sensitive to some prompts. We found overall high robustness along a variety of variations. We noted one important exception with respect to not mentioning to the LLM to consider what it knows so far, a subset of the LLM agents may only return the next trial configuration without any additional thoughts. We found that agents without any "thoughts" were often not stopping their search within the experiment. A possible explanation is that without reflection in the form of written words – which affects the LLM's generative process – the LLM appears to not take into account previous steps it has taken and thereby deviating from more typical human behavior.

## 5 Discussion and Conclusion

Understanding strategizing requires rigorous analyses of both underlying cognitive processes and heuristics used by decision-makers as well as the mechanisms available to organizations to improve outcomes. With our study we show that LLMs can serve as research sandboxes for behavioral strategy scholars, providing a cost-effective environment to both reproduce and extend prior work as well as develop new hypotheses and stress test assumptions in evolving research projects. This new technology is increasingly being explored across disciplines as a tool to inform new theories and hypotheses (Hutson 2023). Similarly, strategy research may gain new insights and enhance understanding of existing empirical work through new approaches (Csaszar et al. 2024). For instance, our extension results using think-aloud output could inspire future scholars to design human lab experiments using specific assumptions and hypotheses concerning how we would expect humans to behave based on the behavior that LLM agents displayed "in the sandbox." These human experiments could then test our finding that forward-looking behavior seems to increase as subjects approach the decision to stop searching and start exploiting prior knowledge.

**5.1 Simulating "human thinking" with LLMs**

Nearly eight decades after Newell and Simon (1959, 1961) began using artificial intelligence to simulate and study human cognitive processes in decision making, LLMs offer intriguing new capabilities that extend and build upon their research tradition in strategic and organizational decision making. The AI models effectively mimic human language patterns, which at least in part can reflect human thinking and reasoning. In the context of behavioral strategy, we have found that LLMs pose great potential in reproducing and extending experimental work. An important insight we have discovered is that LLM simulated agents need to be distinguished from classic computational agents. In formal simulations, variance in results is typically driven by stochastic properties related to the environment (e.g., variation across landscapes) but agent heuristic/behavior is fixed (at least per type of agent). In our LLMs experiments, variation is purely driven by the agent, while we hold the properties of the external environment fixed. Therefore, we see great potential in LLMs complementing formal simulations as they allow scholars to supplement their research assumptions with LLM-generated data. This can help tentatively assess why and when specific cognitive processes and heuristics are invoked.

Another important insight from using LLMs to simulate human thinking is that their outputs can serve as a rich source of "predicted reasoning." In other words, LLMs generate interpretations of rationales, thoughts, or justifications based on their training on vast amounts of human-written content. This training allows the models to infer fundamental semantic relationships, capturing, to some extent, how humans structure their thoughts. Importantly, the thoughts generated by LLMs can effectively capture where, in a specific context, human attention is likely to be focused—based on the model's extensive training. This presents an exciting new avenue for behavioral strategy and decision making, where managerial attention plays a crucial role in understanding differences in firm strategy and performance (Eggers and Kaplan 2009, Joseph and Wilson 2018, Ocasio 2011).

**5.2 Reproducing versus replicating with LLMs**

Our research suggests that the LLM sandbox can support the field's efforts to gain and refine knowledge via replication (Bettis et al. 2016, Fišar et al. 2024). We have demonstrated that LLMs can aid in reproducing and dissecting established findings. However, it is important to distinguish LLM simulated responses from true replication. LLM agent samples, even when yielding similar results, are not replications in its strictest sense, because replication requires the *exact* same research protocol, which is not given if there are no humans involved but artificial intelligence. In addition, the research design in question typically requires adjustments to make it accessible to the LLM. It seems therefore plausible to classify this emerging line of research as reproductions or quasi-replications (Bettis et al. 2016) that offer alternative ways to improve analytical rigor.

      The use of LLM generated thoughts may be particularly useful to drill down on potential rationales underlying human participants' behavior. This is particularly relevant to explore alternative explanations and to narrow down follow-up experiments in the lab when resources and time are constrained. Another important reason for reproducing human samples with LLM

agents can be to address small sample challenges as scaling of LLM outputs is quite cost-efficient (Boussioux et al. 2024). Our findings suggest that researchers may benefit from testing and combining samples from multiple model classes to better reflect variance that occurs in human samples.

**5.3 Future research potential with LLMs**

The approach taken in this study opens avenues for novel research in strategy. In computational simulation studies, one of the main challenges is developing canonical examples that inform assumptions and mechanisms, which are then tested through simulation. LLM experiments, as proposed here, offer a method to offer another robustness-check of these assumptions and possibly inform limitations. Moreover, for experimental studies a growing expectation seems to be to include additional treatments to address methodological concerns (Open Science Collaboration 2015). LLM experiments could help address some of these concerns without necessitating costly and time-consuming additional experimental treatments that may discourage some important research inquiries. For both simulations and experiments, LLM experiments not only offer solutions to these method-specific challenges but also may enhance the validity and generalizability of findings.

A promising avenue for future research involves using LLMs to study and simulate managerial attention in various decision-making scenarios. Research on attention-based decision-making has a long-standing tradition in the Carnegie school, and recent developments have introduced new approaches (e.g., Joseph et al. 2024). LLMs can complement this line of inquiry, as their architecture is specifically designed to understand and interpret human-relevant context by encoding attention scores to words and concepts (Vaswani et al. 2017). Furthermore, by studying attention dynamics between actors—such as through simulations with multiple LLM

agents—scholars can begin to explore organizational dynamics in phenomena related to hierarchy and task divisions (Keum and See 2017, Raveendran et al. 2015).

**5.4 Limitations**

It is important to critically discuss limitations underlying this research. First, we used a single experimental paradigm, the Alien game. While this choice helps connect research traditions in behavioral strategy, further investigation is required with different games and setups to broaden its applicability. Second, because the LLM experiment can only reproduce, not replicate, classical comparisons with statistical methods offer limited insights. Given the particularities of LLMs discussed throughout the paper, a more detailed exploration of how comparisons and calibrations should be performed is necessary. Third, we focused on reproducing our experiment using GPT-4o and GPT-3.5-turbo. Although we tested a few other LLMs, future research could explore more models, as noted by Doshi et al. (2024), particularly in terms of their distinct characteristics, variance handling, and the potential for mixing and matching different models.

Lastly, as LLMs emerge as a frontier in behavioral sciences (Meng 2024), they are often criticized for their lack of understanding, tendency to hallucinate (i.e., provide incorrect or fabricated information), and incomplete consideration of context or data (e.g., Bender et al. 2021). Some argue that "thoughts" produced by LLMs are not equivalent to human thoughts but rather mimic them. While this may not always pose a problem—since simulating human-like responses learned from vast corpora can be useful—more research is needed to understand when and why these models deviate from expected behavior. Additionally, critiques of LLMs have prompted broader concerns about the role of generative AI in research and how to assess the risks associated with their use (Messeri and Crockett 2024). As scholars, we must carefully consider how AI influences our research questions and the choices we make in pursuing them.

While we are confident that AI is already an invaluable addition to scientific research, its implications are not well understood and need to be explored by our community.

**5.5 Conclusion**

We have demonstrated the promising potential of large language models (LLMs) to advance behavioral strategy research in innovative and exciting ways. By reproducing and extending experimental work on strategic search behavior, we illustrate how LLMs can serve as tools for simulating human cognition and decision-making. We do not believe or propose that LLMs should replace traditional computational agents or human subjects. Rather, we argue that LLMs offer an important complementary approach - capturing cognitive nuances that formal models must hard-code, while providing more scalable and controllable data than human experiments alone. We hope our work sparks scholars' interest and encourages the exploration of LLM-based experiments in our field. By investigating this new technology in scientific discourse, behavioral strategy research may advance in novel ways to uncover how cognition affects decision-making.

# REFERENCES

Anvari F, Billinger S, Analytis PP, Franco VR, Marchiori D (2024) Testing the convergent validity, domain generality, and temporal stability of selected measures of people's tendency to explore. *Nat. Commun.* 15(1):7721.

Ashokkumar A, Hewitt L, Ghezae I, Willer R (2024) Predicting Results of Social Science Experiments Using Large Language Models.

Baumann O, Schmidt J, Stieglitz N (2019) Effective search in rugged performance landscapes: A review and outlook. *J. Manag.* 45(1):285–318.

Bender EM, Gebru T, McMillan-Major A, Shmitchell S (2021) On the Dangers of Stochastic Parrots: Can Language Models Be Too Big? 🦜. *Proc. 2021 ACM Conf. Fairness Account. Transpar.* (ACM, Virtual Event Canada), 610–623.

Bergenholtz C, Vuculescu O, Amidi A (2023) Microfoundations of Adaptive Search in Complex Tasks: The Role of Cognitive Abilities and Styles. *Organ. Sci.* 34(6):2043–2063.

Bettis RA, Helfat CE, Shaver JM (2016) The necessity, logic, and forms of replication. *Strateg. Manag. J.* 37(11):2193–2203.

Billinger S, Srikanth K, Stieglitz N, Schumacher TR (2021) Exploration and exploitation in complex search tasks: How feedback influences whether and where human agents search. *Strateg. Manag. J.* 42(2):361–385.

Billinger S, Stieglitz N, Schumacher TR (2014) Search on rugged landscapes: An experimental study. *Organ. Sci.* 25(1):93–108.

Bisbee J, Clinton JD, Dorff C, Kenkel B, Larson JM (2024) Synthetic Replacements for Human Survey Data? The Perils of Large Language Models. *Polit. Anal.*:1–16.

Boussioux L, Lane JN, Zhang M, Jacimovic V, Lakhani KR (2024) The Crowdless Future? Generative AI and Creative Problem-Solving. *Organ. Sci.*:orsc.2023.18430.

Chase WG, Simon HA (1973) Perception in chess. *Cognit. Psychol.* 4(1):55–81.

Csaszar FA, Ketkar H, Kim H (2024) Artificial Intelligence and Strategic Decision-Making: Evidence from Entrepreneurs and Investors. https://www.ssrn.com/abstract=4913363.

Csaszar FA, Levinthal DA (2016) Mental representation and the discovery of new strategies. *Strateg. Manag. J.* 37(10):2031–2049.

Csaszar FA, Siggelkow N (2010) How Much to Copy? Determinants of Effective Imitation Breadth. *Organ. Sci.* 21(3):661–676.

Cyert RM, March JG (1963) *A behavioral theory of the firm* (Blackwell, Oxford).

Doshi AR, Bell JJ, Mirzayev E, Vanneste B (2024) Generative Artificial Intelligence and Evaluating Strategic Decisions. *SSRN Electron. J.*

Eggers JP, Kaplan S (2009) Cognition and Renewal: Comparing CEO and Organizational Effects on Incumbent Adaptation to Technical Change. *Organ. Sci.* 20(2):461–477.

Ericsson KA, Simon HA (1998) How to Study Thinking in Everyday Life: Contrasting Think-Aloud Protocols With Descriptions and Explanations of Thinking. *Mind Cult. Act.* 5(3):178–186.

Farjoun M (2008) Strategy making, novelty and analogical reasoning — commentary on Gavetti, Levinthal, and Rivkin (2005). *Strateg. Manag. J.* 29(9):1001–1016.

Fišar M, Greiner B, Huber C, Katok E, Ozkes AI, and the Management Science Reproducibility Collaboration (2024) Reproducibility in *Management Science*. *Manag. Sci.* 70(3):1343–1356.

Franzen C (2024) Forget GPT-5! OpenAI launches new AI model family o1 claiming PhD-level performance. *VentureBeat*.

Gaessler F, Piezunka H (2023) Training with AI: Evidence from chess computers. *Strateg. Manag. J.* 44(11):2724–2750.

Ganco M, Kapoor R, Lee GK (2020) From Rugged Landscapes to Rugged Ecosystems: Structure of Interdependencies and Firms' Innovative Search. *Acad. Manage. Rev.* 45(3):646–674.

Gavetti G, Greve HR, Levinthal DA, Ocasio W (2012) The Behavioral Theory of the Firm: Assessment and Prospects. *Acad. Manag. Ann.* 6(1):1–40.

Gavetti G, Levinthal D (2000) Looking forward and looking backward: Cognitive and experiential search. *Adm. Sci. Q.* 45(1):113–137.

Gavetti G, Levinthal DA, Rivkin JW (2005) Strategy making in novel and complex worlds: The power of analogy. *Strateg. Manag. J.* 26(8):691–712.

Gavetti G, Rivkin JW (2007) On the origin of strategy: Action and cognition over time. *Organ. Sci.* 18(3):420–439.
Gobet F, Simon HA (1996) Templates in chess memory: A mechanism for recalling several boards. *Cognit. Psychol.* 31(1):1–40.
Hamming RW (1950) Error detecting and error correcting codes. *Bell Syst. Tech. J.* 29(2):147–160.
Heckman J (1979) Sample selection bias as a specification error. *Econometrica*.
Horton JJ (2023) Large Language Models as Simulated Economic Agents: What Can We Learn from Homo Silicus? (January 18) http://arxiv.org/abs/2301.07543.
Joseph J, Laureiro-Martinez D, Nigam A, Ocasio W, Rerup C (2024) Research frontiers on the attention-based view of the firm. *Strateg. Organ.* 22(1):6–17.
Joseph J, Wilson AJ (2018) The growth of the firm: An attention-based view. *Strateg. Manag. J.* 39(6):1779–1800.
Kaplan S (2011) Research in Cognition and Strategy: Reflections on Two Decades of Progress and a Look to the Future: Cognition and Strategy. *J. Manag. Stud.* 48(3):665–695.
Kauffman SA (1993) *The origins of order: Self-organization and selection in evolution* (Oxford University Press, New York).
Keum DD, See KE (2017) The influence of hierarchy on idea generation and selection in the innovation process. *Organ. Sci.* 28(4):653–669.
Klingebiel R (2022) Motivating Innovation: Tunnels vs. Funnels. *Strategy Sci.* 7(4):300–316.
Knudsen T, Srikanth K (2014) Coordinated exploration organizing joint search by multiple specialists to overcome mutual confusion and joint myopia. *Adm. Sci. Q.*:0001839214538021.
Laureiro-Martinez D, Arrieta JP, Brusoni S (2023) Microfoundations of Problem Solving: Attentional Engagement Predicts Problem-Solving Strategies. *Organ. Sci.* 34(6):2207–2230.
Laureiro-Martínez D, Brusoni S (2018) Cognitive flexibility and adaptive decision-making: Evidence from a laboratory study of expert decision makers. *Strateg. Manag. J.* 39(4):1031–1058.
Leiblein MJ, Reuer JJ, Zenger T (2018) What makes a decision strategic? *Strategy Sci.* 3(4):558–573.
Leighton JP (2017) *Using think-aloud interviews and cognitive labs in educational research* (Oxford University Press).
Lenox MJ, Rockart SF, Lewin AY (2006) Interdependency, competition, and the distribution of firm and industry profits. *Manag. Sci.* 52(5):757–772.
Levine SS, Bernard M, Nagel R (2017) Strategic Intelligence: The Cognitive Capability to Anticipate Competitor Behavior. *Strateg. Manag. J.* 38(12):2390–2423.
Levinthal DA (1997) Adaptation on rugged landscapes. *Manag. Sci.* 43(7):934–950.
Levinthal DA (2011) A behavioral approach to strategy—what's the alternative? *Strateg. Manag. J.* 32(13):1517–1523.
Li P, Castelo N, Katona Z, Sarvary M (2024) Frontiers: Determining the Validity of Large Language Models for Automated Perceptual Analysis. *Mark. Sci.* 43(2):254–266.
March JG (1991) Exploration and Exploitation in Organizational Learning. *Organ. Sci.* 2(1):71–87.
March JG, Simon HA (1958) *Organizations* (Wiley, New York).
Mei Q, Xie Y, Yuan W, Jackson MO (2024) A Turing test of whether AI chatbots are behaviorally similar to humans. *Proc. Natl. Acad. Sci.* 121(9):e2313925121.
Meissner P, Poensgen C, Wulf T (2021) How hot cognition can lead us astray: The effect of anger on strategic decision making. *Eur. Manag. J.* 39(4):434–444.
Meng J (2024) AI emerges as the frontier in behavioral science. *Proc. Natl. Acad. Sci.* 121(10):e2401336121.
Messeri L, Crockett MJ (2024) Artificial intelligence and illusions of understanding in scientific research. *Nature* 627(8002):49–58.
Mikolov T, Yih W tau, Zweig G (2013) Linguistic regularities in continuous space word representations. *Proc. 2013 Conf. North Am. Chapter Assoc. Comput. Linguist. Hum. Lang. Technol.* 746–751.
Newell A, Simon HA (1959) *The simulation of human thought* (Rand Corporation Santa Monica, CA, USA).
Newell A, Simon HA (1961) Computer Simulation of Human Thinking: A theory of problem solving expressed as a computer program permits simulation of thinking processes. *Science* 134(3495):2011–2017.
Ocasio W (1997) Towards an Attention-Based View of the Firm. *Strateg. Manag. J.* 18(Summer Special Issue):187–206.

Ocasio W (2011) Attention to Attention. *Organ. Sci.* 22(5):1286–1296.
Open Science Collaboration (2015) Estimating the reproducibility of psychological science. *Science* 349(6251):aac4716.
Piezunka H, Schilke O (2023) The Dual Function of Organizational Structure: Aggregating and Shaping Individuals' Votes. *Organ. Sci.* 34(5):1914–1937.
Porter ME, Siggelkow N (2008) Contextuality Within Activity Systems and Sustainability of Competitive Advantage. *Acad. Manag. Perspect.* 22(2):34–56.
Posen HE, Keil T, Kim S, Meissner FD (2018) Renewing Research on Problemistic Search—A Review and Research Agenda. *Acad. Manag. Ann.* 12(1):208–251.
Raveendran M, Puranam P, Warglien M (2015) Object Salience in the Division of Labor: Experimental Evidence. *Manag. Sci.* 62(7):2110–2128.
Raveendran M, Puranam P, Warglien M (2022) Division of Labor Through Self-Selection. *Organ. Sci.* 33(2):810–830.
Reypens C, Levine SS (2018) Behavior in behavioral strategy: Capturing, measuring, analyzing. *Behav. Strategy Perspect.* (Emerald Publishing Limited), 221–246.
Richter V, Janjic R, Klapper H, Keck S, Reitzig M (2023) Managing exploration in organizations: The effect of superior monitoring on subordinate search behavior. *Strateg. Manag. J.* 44(9):2226–2254.
Rivkin JW (2000) Imitation of complex strategies. *Manag. Sci.* 46(6):824–844.
Siggelkow N, Rivkin JW (2006) When exploration backfires: Unintended consequences of multilevel organizational search. *Acad. Manage. J.* 49(4):779–795.
Simon HA (1947) *Administrative behavior: A study of decision-making processes in administrative organization* (Macmillan).
Steen EV den (2017) A formal theory of strategy. *Manag. Sci.* 63(8):2616–2636.
Strachan JWA, Albergo D, Borghini G, Pansardi O, Scaliti E, Gupta S, Saxena K, et al. (2024) Testing theory of mind in large language models and humans. *Nat. Hum. Behav.* 8(7):1285–1295.
Tracy WM, Markovitch DG, Peters LS, Phani BV, Philip D (2017) Algorithmic representations of managerial search behavior. *Comput. Econ.* 49:343–361.
Vaswani A (2017) Attention is all you need. *Adv. Neural Inf. Process. Syst.*
Vuculescu O (2017) Searching far away from the lamp-post: An agent-based model. *Strateg. Organ.* 15(2):242–263.
Winter SG, Cattani G, Dorsch A (2007) The value of moderate obsession: Insights from a new model of organizational search. *Organ. Sci.* 18(3):403–419.

# FIGURES & TABLES

Figure 1. LLM and Human sample comparison of active search per trial.

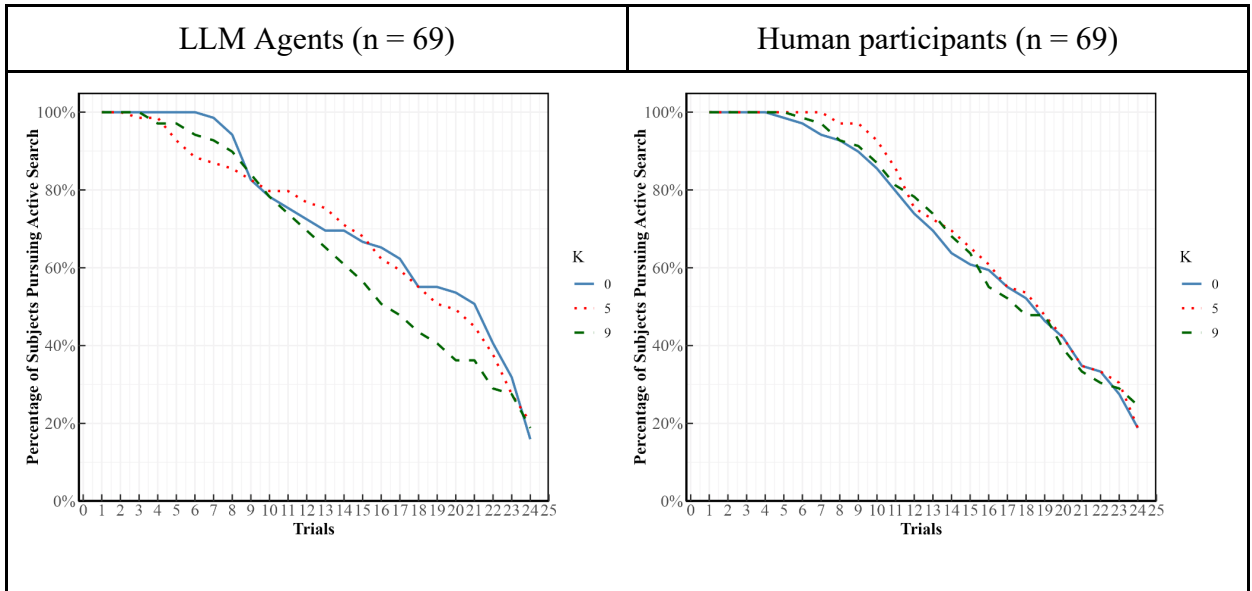

*Note.* Active search refers to an agent or participant no longer making any changes to their tested configuration moving forward.

Figure 2. LLM and Human Comparison of Search Distance by landscape and trial.

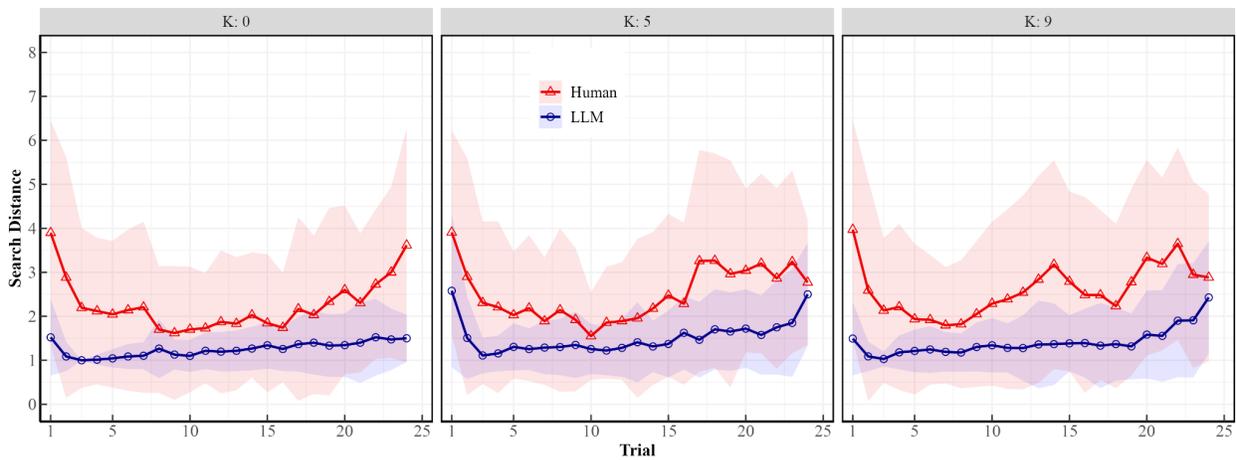

*Note.* Search distance is measured as Hamming distance, that is, the number of bit-wise differences between the highest identified configuration and the configurations tested in a given trial period. The envelopes show the respective standard deviation.

Figure 3. Ratio of LLM forward looking thoughts.

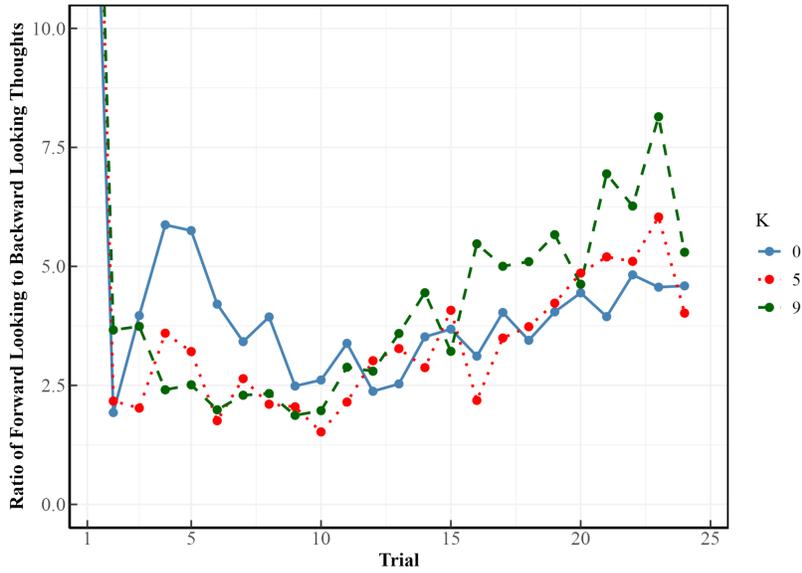

*Note.* The plot shows the means for 300 LLM agents for each landscape complexity.

Figure 4. Mixed LLM Population analysis of search distance compared to human sample.

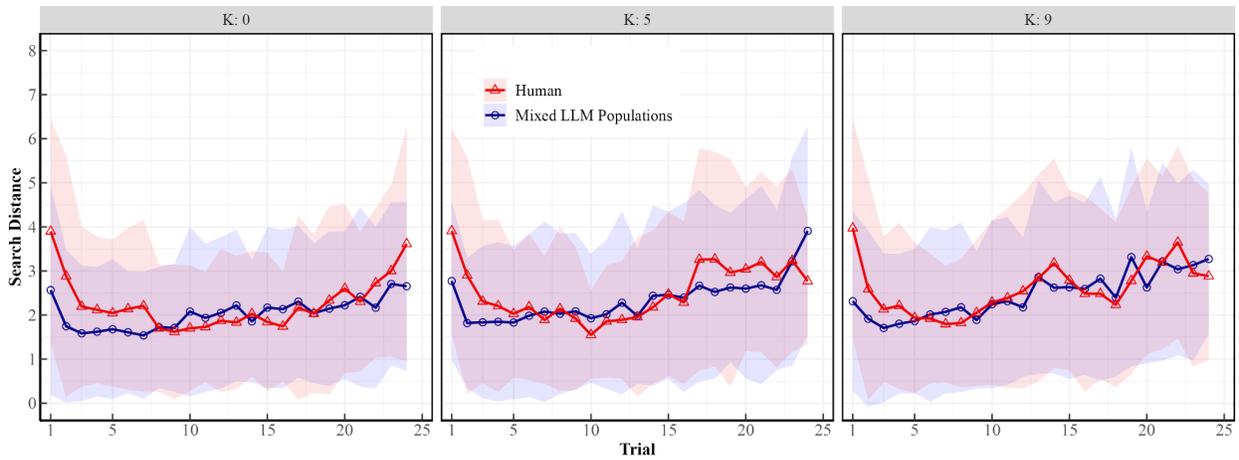

*Note.* Search distance is measured as Hamming distance, that is, the number of bit-wise differences between the highest identified configuration and the configurations tested in a given trial period. The envelopes show the respective standard deviation.

Table 1. Regression for LLM generated think-aloud output.

| Variables | 1st Step Heckman Active Search | | | 2nd Step Heckman Search Distance | | |
|---|---|---|---|---|---|---|
| | Coef. | S.E. | P-Value | Coef. | S.E. | P-Value |
| Attention Breadth | 0.067 | 0.005 | 0.000 | 0.018 | 0.003 | 0.000 |
| Forward Looking Ratio | 0.011 | 0.003 | 0.001 | 0.006 | 0.001 | 0.000 |
| Forward Looking Ratio X Trial | -0.001 | 0.000 | 0.000 | 0.000 | 0.000 | 0.271 |
| Trial | -0.045 | 0.003 | 0.000 | 0.032 | 0.002 | 0.000 |
| Early Feedback | 2.014 | 0.203 | 0.000 | | | |
| Average Feedback | -14.585 | 0.479 | 0.000 | -0.923 | 0.224 | 0.000 |
| Immediate Feedback | 1.704 | 0.200 | 0.000 | 1.075 | 0.073 | 0.000 |
| Reference | 5.671 | 0.301 | 0.000 | -2.193 | 0.138 | 0.000 |
| Prior Search Distance | 1.342 | 0.030 | 0.000 | 0.432 | 0.009 | 0.000 |
| K = 5 | -1.003 | 0.045 | 0.000 | 0.031 | 0.020 | 0.117 |
| K = 9 | -1.722 | 0.061 | 0.000 | -0.112 | 0.027 | 0.000 |
| *Observations* | 21,600 | | | 14,693 | | |
| *Log-Likelihood* | -5703.133 | | | | | |
| *Pseudo R-squared* | | | | 0.547 | | |

Table 2. Overview of regression variables.

| Variable | Explanation |
| --- | --- |
| Attention breadth | Number of distinct symbols (i.e., Greek letters) mentioned in think-aloud output. |
| Forward looking ratio | Ratio of character count of output classified into forward looking text to backward looking text. |
| Trial | Trial number (1 to 24) |
| Early feedback | Highest payoff achieved in the first three trials |
| Average feedback | Average payoff received |
| Immediate feedback | Payoff of the tested configuration in t-1 |
| Reference | Highest payoff received so far |
| Prior search distance | Search distance in trial t-1 |
| K | Landscape complexity (0, 5, 9) |